\documentclass[aip,twocolumn]{revtex4}

\usepackage{amsmath}
\usepackage{amssymb}
\usepackage{stmaryrd}

\usepackage{graphicx}

\usepackage{epsfig}

\def\ov#1{\overline{#1}}

\def\vb#1{\mbox{\boldmath$#1$}}
\def\pd#1#2{\frac{\partial #1}{\partial #2}}

\def\wh#1{\widehat{#1}}
\def\bdot{\,\vb{\cdot}\,}
\def\btimes{\,\vb{\times}\,}

\def\bhat{\wh{{\sf b}}}
\def\cal#1{\mathcal{#1}}

\def\bhat{\wh{{\sf b}}}
\def\exd{{\sf d}}

\newcommand{\bc}{\begin{center}}
\newcommand{\ec}{\end{center}}
\newcommand{\bt}{\begin{tabbing}}
\newcommand{\et}{\end{tabbing}}
\newcommand{\be}{\begin{eqnarray*}}
\newcommand{\ee}{\end{eqnarray*}}
\newcommand{\bs}{\begin{slide}}
\newcommand{\es}{\end{slide}}

\begin{document}

\title{On the validity of the guiding-center approximation \\ in the presence of strong magnetic gradients}

\author{Alain J.~Brizard}
\affiliation{Department of Physics, Saint Michael's College, Colchester, VT 05439, USA}

\date{April 4, 2017}

\begin{abstract}
The motion of a charged particle in a nonuniform straight magnetic field with a constant magnetic-field gradient is solved exactly in terms of elliptic functions. The connection between this problem and the guiding-center approximation is discussed. It is shown that, for this problem, the predictions of higher-order guiding-center theory agree very well with the orbit-averaged particle motion and hold well beyond the standard guiding-center limit $\epsilon \equiv \rho/L \ll 1$, where $\rho$ is the gyromotion length scale and $L$ is the magnetic-field gradient length scale.
\end{abstract}


\maketitle

\section{Introduction} 

The guiding-center dynamics of charged particles moving in a nonuniform magnetic field plays a crucial role in our understanding of magnetically-confined laboratory and space plasmas \cite{Cary_Brizard_2009,Tronko_Brizard_2015}. The standard guiding-center ordering used in deriving these guiding-center equations of motion is based on the dimensionless small parameter $\epsilon \equiv \rho/L \ll 1$, where $\rho$ is the gyromotion length scale and $L$ is the magnetic-field gradient length scale \cite{Note}. Because of its important extensions to the gyrokinetic self-consistent treatment of low-frequency fluctuations in magnetized plasmas \cite{Brizard_Hahm_2007}, it is necessary to gain a full understanding of the validity of the guiding-center approximation, especially when plasma gradients are strong  (see, e.g., Refs.~\cite{Dimits_2012} and \cite{Coury_2016}). For example, in the pedestal region of advanced tokamak plasmas \cite{Coury_2016}, the gradient length scale can be as small as $L \simeq 1-2$ cm, which means that a 10 keV proton confined by a 5 T magnetic field (with a thermal gyroadius $\rho = 2$ mm) is represented by $\epsilon \simeq 0.1-0.2$, which falls well outside the standard guiding-center limit $\epsilon \ll 1$.

In order to investigate the validity of the guiding-center approximation in the presence of strong gradients, we consider the simplest non-trivial problem of a charged particle (with mass $m$ and charge $e$) moving in a straight magnetic field with a uniform magnetic-field gradient:
\begin{equation}
{\bf B}(y) \;=\; B_{0} \left( 1 \;-\frac{}{} y/L\right)\,\wh{\sf z},
\label{eq:B_def}
\end{equation}
where $B_{0}$ is a constant and $L \equiv |\nabla\ln B|^{-1}$ is the constant gradient length scale. 

The Lorentz-force equations for the perpendicular motion in the $(x,y)$-plane are
\begin{eqnarray}
x^{\prime\prime} & = & ( 1 - \epsilon\;y)\;y^{\prime}, \label{eq:x_pp} \\
y^{\prime\prime} & = & -\;( 1 - \epsilon\;y)\;x^{\prime}, \label{eq:y_pp}
\end{eqnarray}
where we introduced the dimensionless coordinates $(x,y) \equiv (x/\rho,y/\rho)$ and a prime denotes a derivative with respect to $\varphi \equiv \Omega\,t$ (with $ \Omega = e B_{0}/mc$). Since the magnetic field is straight $(\bhat = \wh{\sf z})$, we ignore parallel motion along the $z$-axis. We note that these equations satisfy the energy conservation law $x^{\prime 2} + y^{\prime 2} \equiv 2\mu\,B_{0}/(m\rho^{2}\Omega^{2}) = 1$, where $\mu$ denotes the lowest-order magnetic moment.

\begin{figure}
\epsfysize=1.8in
\epsfbox{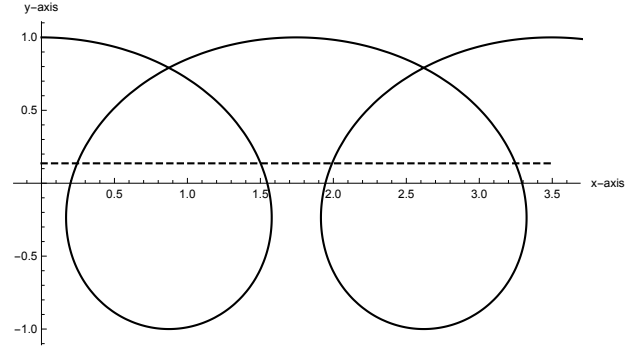}
\caption{Numerical solution of Eqs.~\eqref{eq:x_pp}-\eqref{eq:y_pp}, subject to the initial conditions \eqref{eq:xy_init}, for $\epsilon = 0.5$. The dashed horizontal line denotes the orbit average
\eqref{eq:y_orbit} of the particle's $y$ position.}
\label{fig:num_sol}
\end{figure}

In the present paper, we will study the solution for Eqs.~\eqref{eq:x_pp}-\eqref{eq:y_pp} for $0 \leq \epsilon \leq 1$ (instead of the standard guiding-center limit $\epsilon \ll 1$ \cite{Cary_Brizard_2009}), subject to the initial conditions
\begin{equation}
\left(x_{0}, x_{0}^{\prime};\frac{}{} y_{0}, y_{0}^{\prime}\right) \;=\; (0, 1;\, 1,0).
\label{eq:xy_init}
\end{equation}
Figure \ref{fig:num_sol} shows the numerical solution of Eqs.~\eqref{eq:x_pp}-\eqref{eq:y_pp},  subject to the initial conditions \eqref{eq:xy_init}, for the case $\epsilon = 0.5$ (i.e., the gyromotion length scale $\rho = 
L/2$ is half of the magnetic-gradient length scale L). Here, we clearly see the standard grad-B drift motion along the $x$-axis due to the magnetic-field gradient (which is exaggerated, here, by choosing $\epsilon = 0.5$).

In Fig.~\ref{fig:num_sol}, we also note that, while the transverse motion along the $y$-axis is periodic in the range $-1 \leq y \leq 1$, the average $y$-position is not zero for $\epsilon \neq 0$. Here, the orbital average (dashed horizontal line) of the particle's $y$ position is evaluated as
\begin{equation}
\ov{y}(\epsilon) \;\equiv\; \frac{1}{4\;{\sf K}(\epsilon^{2})}\;\int_{0}^{4\;{\sf K}(\epsilon^{2})}\; y(\varphi,\epsilon)\; d\varphi,
\label{eq:y_orbit}
\end{equation}
where the complete elliptic integral of the first kind ${\sf K}(m)$ is defined as \cite{Mathematica}
\begin{equation}
{\sf K}(m) \;\equiv\; \int_{0}^{\pi/2}\frac{d\varphi}{\sqrt{1 - m\,\sin^{2}\varphi}}.
\label{eq:K_def}
\end{equation}
In the uniform limit (i.e., $\epsilon = 0$), we find $\ov{y}(0) = 0$, as expected for the trigonometric solution $y(\varphi,0) = \cos\varphi$, with period $4{\sf K}(0) = 2\pi$. 

The purpose of the present work is to compare the orbit-averaged properties of the exact analytical solution of Eqs.~\eqref{eq:x_pp}-\eqref{eq:y_pp} for the perpendicular motion in the $(x,y)$-plane, which will be explicitly expressed in terms of Jacobi elliptic functions \cite{NIST_Jacobi} and elliptic integrals, with the predictions of higher-order guiding-center theory \cite{Tronko_Brizard_2015}. With these exact orbit averages, we will show that the results of guiding-center theory are valid well outside the guiding-center limit $\epsilon \ll 1$, which is consistent with Mynick's work \cite{Mynick_1979,Mynick_1980}, where guiding-center theory was extended to $\epsilon$ comparable to unity within the context of bounce-center theory. 

We note that the Jacobi elliptic functions have also recently been used in the description of guiding-center and bounce-center dynamics in axisymmetric tokamak geometry \cite{Brizard_2011,Brizard_2014,Duthoit_2014}. They have also been used in obtaining exact analytical solutions for charged-particle motion in nonuniform magnetic fields represented by power laws \cite{Headland_Seymour_1975,Repko_1991}, which includes the case of particle orbits crossing a line where $B = 0$ (case 3 in Ref.~\cite{Headland_Seymour_1975}), and stationary current distributions \cite{Essen_Nordmark_2016}. The case of a discontinuous magnetic field (with a gradient length scale $L = 0$) is also solved exactly in Ref.~\cite{Dodin_Fisch_2001}.

\section{Guiding-center Approximation}  

We begin by deriving the lowest-order guiding-center results to be quoted in the remainder of this paper. Because magnetic-curvature and parallel-gradient effects vanish for the magnetic field \eqref{eq:B_def}, we ignore the parallel motion along field lines. 

Two features associated with the averaged properties of the particle motion are seen in Fig.~\ref{fig:num_sol}: an averaged displacement $\ov{y}(\epsilon)$ along the $y$-axis (in the opposite direction to the magnetic-field gradient) and a drift motion $v_{\rm D}(\epsilon) \equiv x(4{\sf K},\epsilon)/(4{\sf K})$ along the $x$-axis. Both features can easily be explained by higher-order guiding-center theory \cite{Cary_Brizard_2009,Tronko_Brizard_2015}.

First, for the guiding-center approximation of the averaged transverse displacement $\ov{y}(\epsilon)$, we compute the guiding-center averaged particle position $\langle {\bf x}\rangle_{\rm gc} \equiv 
\langle{\sf T}_{\rm gc}{\bf X}\rangle_{\rm gc}$ according to high-order guiding-center theory \cite{Tronko_Brizard_2015}, where
\[ {\sf T}_{\rm gc}{\bf X} \;\equiv\; {\bf X} + G_{1}^{\bf X} + G_{2}^{\bf X} + \frac{1}{2}\,{\sf G}_{1}\cdot\exd G_{1}^{\bf X} + \cdots \]
is defined as the guiding-center pull-back of the guiding-center position ${\bf X} = (X, Y)$ to particle phase space, where the phase-space vector fields $({\sf G}_{1}, {\sf G}_{2},\cdots)$ generate the guiding-center transformation \cite{Tronko_Brizard_2015}. From this expression, we compute the guiding-center averaged-particle displacement
\begin{eqnarray}
\langle{\bf x}\rangle_{\rm gc} \;-\; {\bf X} & = & \langle G_{2}^{\bf X}\rangle_{\rm gc}  \;-\; \frac{1}{2}\,\left\langle{\sf G}_{1}\cdot\exd\vb{\rho}_{0}\right\rangle_{\rm gc}  \nonumber \\
 & = & \frac{1}{2}\;\frac{\mu B_{0}}{m\Omega^{2}}\;\nabla\ln B \;-\; \frac{\mu B_{0}}{m\Omega^{2}}\;\nabla\ln B \nonumber \\
  & = & -\;\frac{\mu B_{0}}{2\,m\Omega^{2}}\;\nabla\ln B,
\label{eq:gc_shift}
\end{eqnarray}
where the gyroradius vector $\vb{\rho}_{0}(Y,\mu,\theta) \equiv -\,G_{1}^{\bf X}$ depends on the gyroangle $\theta$, the guiding-center magnetic moment $\mu$, and the guiding-center coordinate $Y$ (because its magnitude depends on the strength of the magnetic field), and we used $\langle\vb{\rho}_{0}\rangle_{\rm gc} = 0$ in Eq.~\eqref{eq:gc_shift}. From Eq.~\eqref{eq:gc_shift}, we see that the guiding-center-averaged $x$-position of the particle $\langle x \rangle_{\rm gc} = X$ is equal to the guiding-center position $X$. On the other hand, because of the magnetic-field gradient (with $\nabla\ln B = -\,\wh{\sf y}/L$), the normalized guiding-center-averaged particle displacement along the $y$-axis is 
\begin{equation}
\frac{1}{\rho}\,(\langle y\rangle_{\rm gc} - Y) = \wh{\sf y}\bdot\left(-\;\frac{\mu B_{0}}{2\,m\rho\Omega^{2}}\;\nabla\ln B\right) = \frac{\epsilon}{4}.
\label{eq:gc_displacement}
\end{equation}
This guiding-center prediction is, in fact, quite close (see Fig.~\ref{fig:yave}) to the orbit-averaged position \eqref{eq:y_orbit} shown by the solid horizontal line in Fig.~\ref{fig:num_sol}. 

Next, the dimensionless guiding-center drift velocity $\dot{\bf X}/(\rho\Omega)$ in a straight magnetic field with a constant gradient (normalized to the perpendicular particle velocity scale $\rho\Omega$) is 
\cite{Cary_Brizard_2009}
\begin{equation}
\bhat\btimes\left(\frac{\mu\,B_{0}}{m\rho\Omega^{2}}\,\nabla\ln B\right) \;=\; \wh{\sf z}\btimes\left( -\,\frac{\epsilon}{2}\,\wh{\sf y}\right) \;=\; \frac{\epsilon}{2}\,\wh{\sf x}.
\label{eq:vdrift_gc}
\end{equation}
This guiding-center result is clearly seen as the orbit-averaged drift motion along the $x$-axis in Fig.~\ref{fig:num_sol}. The guiding-center prediction \eqref{eq:vdrift_gc} is also quite close to the orbit-averaged drift velocity along the $x$-axis (see Fig.~\ref{fig:vdrift_gc}). We note that the case of a discontinuous magnetic field discussed in Ref.~\cite{Dodin_Fisch_2001} exhibits a drift motion that can be calculated exactly along lines similar to the derivation of Eq.~\eqref{eq:vdrift_gc}.

While these guiding-center predictions \eqref{eq:gc_displacement}-\eqref{eq:vdrift_gc} are derived in the standard guiding-center limit $\epsilon \ll 1$ (i.e., when the magnetic-gradient length scale $L$ is assumed to be much longer than the gyromotion length scale $\rho$), we will show that these guiding-center predictions are, in fact, also valid well outside the limit $\epsilon \ll 1$. 

\section{Exact Transverse Motion}  

We now derive the exact solution for the transverse motion along the $y$-axis, which is expressed in terms of the Jacobi elliptic functions \cite{NIST_Jacobi}. With this solution, we will obtain an explicit expression
for the orbit average \eqref{eq:y_orbit}, which can then be compared with the guiding-center predictions \eqref{eq:gc_displacement}.

\subsection{Exact solution}

We begin by noting that, since the right side of Eq.~\eqref{eq:x_pp} can be expressed as an exact derivative with respect to $\varphi$, we integrate it to obtain
\begin{equation}
x^{\prime} \;=\; \frac{\epsilon}{2} \left( 1 \;-\frac{}{} y^{2}\right) \;+\; y,
\label{eq:x_prime}
\end{equation}
where we used the initial conditions \eqref{eq:xy_init}. By inserting Eq.~\eqref{eq:x_prime} into Eq.~\eqref{eq:y_pp}, we obtain the nonlinear second-order ODE
\[ y^{\prime\prime} = -\;\frac{\epsilon}{2} - \left(1 \;-\; \frac{\epsilon^{2}}{2}\right) y + \frac{3}{2}\,\epsilon\;y^{2} -\frac{\epsilon^{2}}{2}\;y^{3}. \]
Next, we multiply this equation by $2\,y^{\prime}$, and, then using the initial conditions \eqref{eq:xy_init}, we integrate it to finally obtain
\begin{eqnarray}
(y^{\prime})^{2} & = & -\;\epsilon\,(y - 1) \;-\; \left(1 \;-\; \frac{\epsilon^{2}}{2}\right) (y^{2} - 1) \nonumber \\
 &  &+\; \epsilon\;(y^{3} - 1) \;-\; \frac{\epsilon^{2}}{4}\;(y^{4} - 1).
 \label{eq:y_dot_squared}
\end{eqnarray}
We will now seek an exact solution $y(\varphi,\epsilon)$ of Eq.~\eqref{eq:y_dot_squared} for the transverse motion in terms of the Jacobi elliptic functions. 

\begin{figure}
\epsfysize=1.8in
\epsfbox{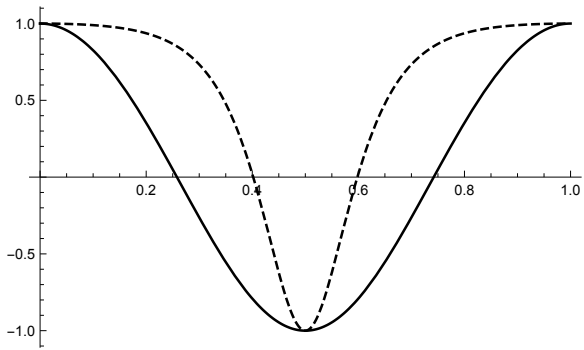}
\caption{Plots of $y(2\omega_{1}\,s, \epsilon)$ in the range $0 \leq s \leq 1$ for the elliptic solutions \eqref{eq:y_J} for $\epsilon = 0.1$ (solid) and $\epsilon = 0.99$ (dashed).}
\label{fig:y_sol}
\end{figure}

For this purpose, we transform Eq.~\eqref{eq:y_dot_squared} by writing
\[ y \;=\; \left(\frac{2}{\epsilon} - 1\right) - \frac{2}{\epsilon} \left(\frac{1 - \epsilon}{1 - \epsilon\,w^{2}}\right), \]
where $w(\varphi,\epsilon)$ is a solution of the differential equation $(w^{\prime})^{2} = \frac{1}{4}\,(1 - w^{2})\,(1 - \epsilon^{2}\,w^{2})$. By using the initial condition $w(0,\epsilon) = 0$, the solution
$w(\varphi,\epsilon) = {\rm sn}(\varphi/2|\epsilon^{2})$ is expressed in terms of the doubly-periodic Jacobi elliptic function ${\rm sn}(z|m)$, with a real period of $4{\sf K}(m)$ and an imaginary period $2i\,{\sf K}(1-m)$. Hence, the transverse motion 
\begin{equation}
y(\varphi,\epsilon) \;=\; \left(\frac{2}{\epsilon} - 1\right) - \frac{2}{\epsilon} \left[\frac{1 - \epsilon}{1 - \epsilon\,{\rm sn}^{2}(\varphi/2|\epsilon^{2})}\right]
\label{eq:y_J}
\end{equation}
corresponds exactly to the numerical solution shown in Fig.~\ref{fig:num_sol}, which is periodic in the range $-1 \leq y \leq 1$ with a period of $4\,{\sf K}(\epsilon^{2})$, as used in Eq.~\eqref{eq:y_orbit}. Note that at the half-period $2\,{\sf K}$, we have the exact result $y(2\,{\sf K},\epsilon) =  -1$, which holds for all values of $\epsilon$. This exact solution is shown in Fig.~\ref{fig:y_sol} for $\epsilon = 0.1$ (solid) and $\epsilon = 0.99$ (dashed). We note that, as $\epsilon$ increases toward 1, the solution spends an increasingly larger portion of its orbit in the range $y > 0$, although it must still satisfy $y(2\,{\sf K},\epsilon) = -1$. This feature explains why the orbit average \eqref{eq:y_orbit} is positive for finite magnetic gradients $\epsilon > 0$.

\subsection{Averaged transverse motion}

We can now use the Jacobi solution \eqref{eq:y_J} to obtain an explicit expression for the orbit average \eqref{eq:y_orbit}. For this purpose, we introduce the definition
\begin{equation}
\int_{0}^{s}\frac{du}{1 - \epsilon\;{\rm sn}^{2}(u|\epsilon^{2})} \;=\; \Pi\left(\epsilon,{\rm am}(s|\epsilon^{2})\;|\;\epsilon^{2}\right),
\end{equation}
expressed in terms of the incomplete elliptic integral of the third kind, where ${\rm am}(s|m) \equiv \arcsin[{\rm sn}(s|m)]$ is the Jacobi amplitude function \cite{NIST_Jacobi}. Hence, we compute the orbit average \eqref{eq:y_orbit} of Eq.~\eqref{eq:y_J}:
\begin{eqnarray}
\ov{y}(\epsilon) & = & \frac{1}{2\omega_{1}}\int_{0}^{2\omega_{1}} y(\varphi,\epsilon)\,d\varphi \;=\; \frac{1}{4{\sf K}}\int_{0}^{4{\sf K}} y(\varphi,\epsilon)\,d\varphi\nonumber \\
 & = & \left(\frac{2}{\epsilon} - 1\right) - \left(\frac{1}{\epsilon} - 1\right) \frac{\Pi\left(\epsilon,\pi\;|\;\epsilon^{2}\right)}{{\sf K}(\epsilon^{2})} \nonumber \\
  & = & \frac{1}{\epsilon}\left(1 \;-\; \frac{\pi}{2{\sf K}(\epsilon^{2})}\right),
 \label{eq:yave_orbit}
\end{eqnarray}
where we used ${\rm am}(2{\sf K}|\epsilon^{2}) = \pi$ and
\[  \frac{\Pi\left(\epsilon,\pi\;|\;\epsilon^{2}\right)}{{\sf K}(\epsilon^{2})} \;=\; 1 \;+\; \frac{\pi}{2(1 - \epsilon)\,{\sf K}(\epsilon^{2})}. \]

\begin{figure}
\epsfysize=1.8in
\epsfbox{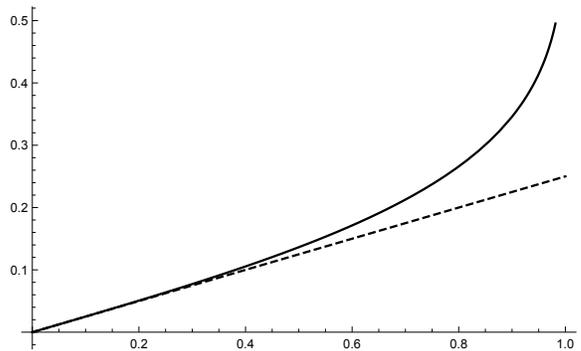}
\caption{Plots of the orbit average $\ov{y}(\epsilon)$ (solid curve) and the lowest-order guiding-center approximation $\epsilon/4$ (dashed line) in the range $0 \leq \epsilon \leq 1$.}
\label{fig:yave}
\end{figure}

The plots of Eq.~\eqref{eq:yave_orbit} (solid curve) and the lowest-order guiding-center approximation $\epsilon/4$ (dashed curve), given by Eq.~\eqref{eq:gc_displacement}, are shown in Fig.~\ref{fig:yave}. We note that the lowest-order guiding-center approximation \eqref{eq:gc_displacement} for the averaged particle $y$-position is remarkably close to the orbit averaged position \eqref{eq:yave_orbit} for values of 
$\epsilon$ well beyond the guiding-center limit $\epsilon \ll 1$. In Fig.~\ref{fig:num_sol}, for example, we find $\ov{y}(0.5) = 0.136...$, and the guiding-center result $\epsilon/4 = 0.125$, which is just 8\% below the orbit-averaged value $\ov{y}(\epsilon)$.  

Lastly, we note that an alternative way to write the transverse solution \eqref{eq:y_J} involves the gyroangle $\theta(\varphi,\epsilon)$, which satisfies the differential equation $\theta^{\prime} = 1 - \epsilon\,
y(\varphi,\epsilon)$. Using Eq.~\eqref{eq:y_J}, its solution is expressed as
\begin{equation}
\theta(\varphi,\epsilon) \;=\; \frac{\pi\,\varphi}{2\,{\sf K}(\epsilon^{2})} \;+\; \Delta\theta(\varphi,\epsilon),
\label{eq:theta}
\end{equation}
where the periodic function 
\begin{equation} 
\Delta\theta(\varphi,\epsilon) \;=\; (1 - \epsilon) \left[ 4\,\Pi\left(\epsilon,{\rm am}(\varphi/2)\right) \;-\;  \frac{\Pi\left(\epsilon,\pi\right)}{{\sf K}(\epsilon^{2})} \;\varphi \right]
\label{eq:Delta_theta}
\end{equation}
vanishes at $\varphi = 0, 2{\sf K}$, and $4{\sf K}$. The transverse solution \eqref{eq:y_J} can, therefore, be expressed as
\begin{equation}
y(\varphi,\epsilon) \;=\; \frac{1}{\epsilon}\left(1 \;-\; \frac{\pi}{2{\sf K}(\epsilon^{2})}\right) \;-\; \frac{1}{\epsilon}\;\pd{\Delta\theta}{\varphi},
\label{eq:y_theta}
\end{equation}
where the orbit average of the second term is explicitly zero because of the periodicity of Eq.~\eqref{eq:Delta_theta}.

\section{Drift Motion}  

\begin{figure}
\epsfysize=1.8in
\epsfbox{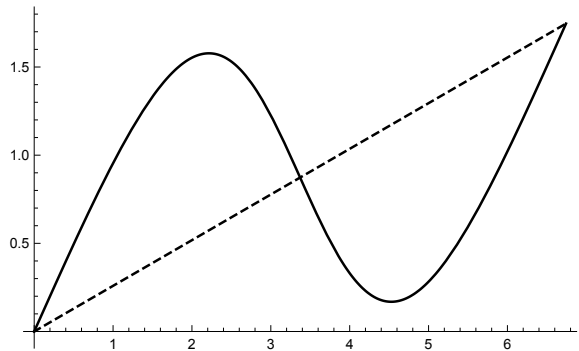}
\caption{Plot of Eq.~\eqref{eq:x_J} in the range $0 \leq \varphi \leq 4\,{\sf K}(\epsilon^{2})$ for the case $\epsilon = 0.5$. The dashed line represents the drift motion along the $x$-axis.}
\label{fig:x_J}
\end{figure}

The solution for $x(\varphi,\epsilon)$ can now be obtained by integrating Eq.~\eqref{eq:x_prime}:
\begin{equation}
x(\varphi,\epsilon) \;=\; \int_{0}^{\varphi} \left[\frac{\epsilon}{2} \left( 1 \;-\frac{}{} y^{2}(t,\epsilon)\right) \;+\; y(t,\epsilon) \right] dt,
\label{eq:x_sol}
\end{equation}
subject to the initial conditions \eqref{eq:xy_init}. By inserting Eq.~\eqref{eq:y_J}, we find the solution
\begin{eqnarray}
x(\varphi,\epsilon)  & = & \frac{\varphi}{\epsilon} \left(1 \;-\; \frac{{\sf E}(\epsilon^{2})}{{\sf K}(\epsilon^{2})}\right) \;-\; \frac{2}{\epsilon} {\cal Z}\left({\rm am}(\varphi/2|\epsilon^{2})\frac{}{}|\;\epsilon^{2}\right) \nonumber \\
  &  &+\; \frac{2\;{\rm sn}(\varphi/2|\epsilon^{2})\,{\rm cn}(\varphi/2|\epsilon^{2})\,{\rm dn}(\varphi/2|\epsilon^{2})}{1 \;-\; \epsilon\,{\rm sn}^{2}(\varphi/2|\epsilon^{2})},
\label{eq:x_J}
\end{eqnarray}
which is expressed in terms of the Jacobi elliptic functions $({\rm sn}, {\rm cn}, {\rm dn})$, the complete elliptic integral of the second kind, ${\sf E}(\epsilon^{2})$, and the Jacobi zeta function \cite{NIST_Jacobi}
\[ {\cal Z}\left({\rm am}(\varphi/2|\epsilon^{2})\frac{}{}|\;\epsilon^{2}\right)  \;\equiv\; \pd{}{\varphi}\ln\left[\vartheta_{4}^{2}(\zeta,q)\right], \]
which is expressed in terms of the logarithmic derivative of the elliptic theta function $\vartheta_{4}(\zeta,q)$ \cite{NIST_Theta}, where $\zeta = \pi\varphi/(4 {\sf K}(\epsilon^{2}))$ and $q(\epsilon) = \exp[-\pi\;{\sf K}(1-\epsilon^{2})/{\sf K}(\epsilon^{2})]$. Since the last term in Eq.~\eqref{eq:x_J} can also be expressed as a logarithmic derivative
\[ \frac{2\;{\rm sn}\,{\rm cn}\,{\rm dn}}{1 \;-\; \epsilon\,{\rm sn}^{2}} \;=\; -\;\frac{2}{\epsilon}\pd{}{\varphi}\ln\left[1 - \epsilon\,{\rm sn}^{2}(\varphi/2|\epsilon^{2})\right], \]
we use the identity \cite{NIST_Theta}
\[ 1 - \epsilon\,{\rm sn}^{2}(\varphi/2|\epsilon^{2}) \;\equiv\; 1 - \vartheta_{1}^{2}(\zeta,q)/\vartheta_{4}^{2}(\zeta,q), \]
so that Eq.~\eqref{eq:x_J} can also be expressed as
\begin{eqnarray}
x(\varphi,\epsilon) & = & \frac{\varphi}{\epsilon} \left(1 \;-\; \frac{{\sf E}(\epsilon^{2})}{{\sf K}(\epsilon^{2})}\right) 
\label{eq:x_theta} \\
 &  &- \frac{\pi}{2\epsilon\,{\sf K}(\epsilon^{2})} \left(\pd{}{\zeta}\ln\left[\vartheta_{4}^{2}(\zeta,q) - \vartheta_{1}^{2}(\zeta,q)\right]\right),
\nonumber
\end{eqnarray}
where the second term is a periodic function of $\zeta$. Once again, this solution, which is shown in Fig.~\ref{fig:x_J}, agrees exactly with the numerical solution shown in Fig.~\ref{fig:num_sol}. It is interesting to note that an expression similar to the second term in Eq.~\eqref{eq:x_theta} has previously appeared in Ref.~\cite{Brizard_2014}, where the generating functions for the canonical transformation for trapped/passing guiding-center orbits in axisymmetric tokamak geometry were derived.

\begin{figure}
\epsfysize=1.8in
\epsfbox{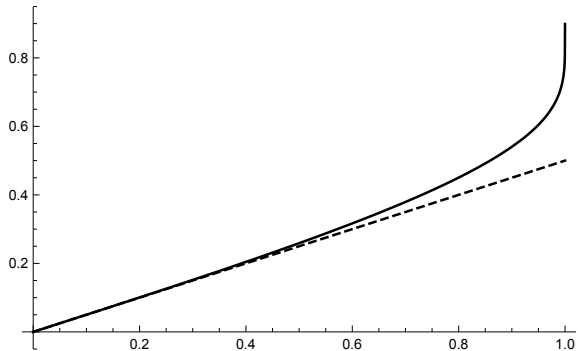}
\caption{Plots of the orbit-averaged particle drift velocity \eqref{eq:v_drift} (solid curve) and the lowest-order guiding-center approximation $\epsilon/2$ (dashed curve) in the range $0 \leq \epsilon \leq 1$.}
\label{fig:vdrift_gc}
\end{figure}

The dashed line  in Fig.~\ref{fig:x_J} represents the drift motion along the $x$-axis, with a slope equal to the orbit-averaged particle drift velocity obtained from Eq.~\eqref{eq:x_theta} :
\begin{equation}
v_{\rm D}(\epsilon) \equiv \frac{x(4{\sf K},\epsilon)}{4\,{\sf K}} \;=\; \frac{1}{\epsilon} \left(1 \;-\; \frac{{\sf E}(\epsilon^{2})}{{\sf K}(\epsilon^{2})}\right).
 \label{eq:v_drift}
\end{equation}
Equation \eqref{eq:v_drift} and the lowest-order guiding-center prediction $\epsilon/2$, derived in Eq.~\eqref{eq:vdrift_gc}, are shown in Fig.~\ref{fig:vdrift_gc}. Once again, the lowest-order guiding-center prediction \eqref{eq:vdrift_gc} yields an excellent agreement with the exact result \eqref{eq:v_drift} well outside the standard guiding-center limit $\epsilon \ll 1$. For example, for the case $\epsilon = 0.5$, the guiding-center prediction $\epsilon/2$ yields $0.5/2 = 0.25$, which is just 3\% below the orbit-averaged particle drift velocity $v_{\rm D}(0.5) = 0.259...$

\section{Summary}  

By solving exactly the motion of a charged particle in a straight magnetic field \eqref{eq:B_def} with a constant gradient, we have been able to investigate the validity of the guiding-center approximation. In the present work, we have shown that the guiding-center predictions \eqref{eq:gc_displacement} and \eqref{eq:vdrift_gc} agree very well with the orbit-averaged particle displacement \eqref{eq:yave_orbit} and the orbit-averaged particle drift velocity \eqref{eq:v_drift} for values of $\epsilon$ well outside the standard guiding-center limit $\epsilon \ll 1$. 

These results can, thus, be used to justify the applications of gyrokinetic theory for magnetized plasmas with strong gradients \cite{Dimits_2012}. For example, in the pedestal region of advanced tokamak plasmas \cite{Coury_2016}, where $\epsilon \simeq 0.2$, the guiding-center predictions \eqref{eq:gc_displacement} and \eqref{eq:vdrift_gc} are 1.3\% and 0.5\% below the orbit-averaged particle displacement \eqref{eq:yave_orbit} and the orbit-averaged particle drift velocity \eqref{eq:v_drift}, respectively.

\acknowledgments

The author wishes to thank H.~Ess\'{e}n for pointing out Refs.~\cite{Headland_Seymour_1975} and \cite{Essen_Nordmark_2016}. This work was supported by the U.S.~Dept.~of Energy under contract No.~DE-SC0014032.


\begin{thebibliography}{99}
\bibitem{Note} For example, the thermal gyroadius of a 10 keV proton confined by a 5 T magnetic field (with a gradient length scale $L =$ 1 m) is $\rho =$ 2 mm and, thus, the guiding-center dimensionless parameter $\epsilon \equiv \rho/L = 2\,\times 10^{-3}$ is well within the guiding-center limit $\epsilon \ll 1$.
\bibitem{Mathematica} Throughout the text, we use the {\sf Mathematica} notation for the complete and incomplete elliptic integrals as well as the Jacobi elliptic functions. For example, ${\sf K}(m)$ in {\sf Mathematica} notation is ${\sf K}(k)$ in standard notation with $m = k^{2}$.
\end{thebibliography}
\end{document}